\documentclass[aps,showpacs,prb,superscriptaddress,floatfix,twocolumn,citeautoscript]{revtex4-1}
\usepackage{graphicx}
\usepackage{amsmath}
\usepackage{amssymb}
\usepackage{bm}
\usepackage{dcolumn}
\usepackage{subfigure} 
\usepackage{hyperref}
\usepackage[usenames]{color}
\usepackage{booktabs}
\usepackage{multirow}
\usepackage{siunitx}

\usepackage{wrapfig}
\usepackage{lipsum}
\usepackage{setspace}
\hypersetup{pdfborder=0 0 0,colorlinks=true,citecolor=blue,linkcolor=blue}
\input epsf
\newcommand{\BN}{\textit{h}-BN}

\begin{document}

\title{Anisotropic charge density wave in electron-hole double monolayers: Applied to phosphorene} 
\author{S. Saberi-Pouya}
\email{samira.saberipouya@uantwerpen.be}
\affiliation{Department of Physics, Shahid Beheshti University, G. C., Evin, Tehran 1983969411, Iran}
\affiliation{Department of Physics, University of Antwerp, Groenenborgerlaan 171, B-2020 Antwerpen, Belgium}
\author{M. Zarenia}
\affiliation{Department of Physics, University of Antwerp, Groenenborgerlaan 171, B-2020 Antwerpen, Belgium}
\author{T. Vazifehshenas}
\affiliation{Department of Physics, Shahid Beheshti University, G. C., Evin, Tehran 1983969411, Iran}
\author{F. M. Peeters}
\affiliation{Department of Physics, University of Antwerp, Groenenborgerlaan 171, B-2020 Antwerpen, Belgium}

\begin{abstract}	
	
	The possibility of an inhomogeneous charge density wave phase is investigated in a system of two coupled electron and hole monolayers separated by a hexagonal boron nitride insulating layer. The charge density wave state is induced through the assumption of negative compressibility of electron/hole  gases in a Coulomb drag configuration between the electron and hole sheets. Under equilibrium conditions, we derive  analytical expressions for the density oscillation along the zigzag and armchair directions. We find that the density modulation not only depends on the sign of the compressibility but also on the anisotropy of the low energy bands.  Our results are applicable to any two dimensional system with anisotropic parabolic bands, characterized by different effective masses. For equal effective masses, i.e., isotropic energy bands, our results agree with Hrolak \textit{et al.} \cite{PhysRevB.96.075422}. Our numerical results are applied to phosphorene. 
\end{abstract}

\date{\today}
\pacs{71.35.-y, 73.21.-b, 74.78.Fk}
\maketitle
%
\section{Introduction}

Phosphorene i.e. a monolayer of black phosphorus(BP), has been investigated theoretically and experimentally, which has attracted a lot of interest due to its highly anisotropic electronic and optical properties \cite{2053-1583-4-2-025064,PhysRevB.92.081408,Samira:drag2016,zhu2016black,2053-1583-4-2-025071,Samira:conductivity2017}. Phosphorene has a puckered honeycomb structure with each phosphorus atom covalently bonded with three adjacent atoms. Unlike graphene, phosphorene is not perfectly flat and forms a puckered surface due to the $sp^{3}$ hybridization of the 3$s$ and 3$p$ atomic orbitals. The band gap of phosphorene is direct and can be tuned from 2 \text{eV} to the visible spectrum by applying a perpendicular electric field \cite{Rodin:prl2014,PhysRevB.90.205421}. More importantly, phosphorene exhibits a strong in-plane anisotropy, which is absent in graphene and in most transition metal dichalcogenides (TMDs) \cite{Phosphorene:review}.  BP has recently  emerged as a new two dimensional (2D) material for high performance electronic and optoelectronic devices because of its high mobility, tunable mid-infrared band gap, anisotropic electronic properties and possible superfluidity\cite{PhysRevB.97.174503}. The anisotropic response of BP was also determined dynamically with a 200-fs time resolution by the convolution of pump-probe pulses \cite{Photoexcitation:nanol15}. The exciton binding energy for direct excitons in phosphorene was obtained experimentally by polarization-resolved photoluminescence measurements at room temperature \cite{excitonsBP:nnano15}. 

In comparison to graphene, phosphorene is chemically reactive and tends to form strong bonds with the surface of substrates, which leads to structural changes  \cite{C4CC05752J,2053-1591-3-2-025013,PhysRevB.96.115402}. Naturally, chemically stable 2D material systems, such as graphene and hexagonal boron nitride (\BN), may be used to protect the fragile, low chemical stability of phosphorene \cite{Bokdam:nanol11}. Encapsulation of few-layer phosphorene by \BN \ sheets has resulted in new ultraclean heterostructures which could be ideal anisotropic 2D systems with high mobility and a possible negative compressibility of electron/hole gas ($ \kappa^{-1}=n^2 \partial \mu / \partial n $), where $n$ is the density and
$\mu$ is the chemical potential of the interacting system\cite{PhysRevB.93.035455,PhysRevB.90.035404, PhysRevB.50.1760,PhysRevB.96.075422}. A negative compressibility results from electron-electron interactions, in which the exchange and correlation energies lower the chemical potential as the electron density decreases. This effect has been observed to enhance the capacitance of semiconductor 2D electronic systems by a few percent above the expected geometric capacitance \cite{Capacitance2011} and is experimental evidence for the formation of a charge-density-wave (CDW) phase \cite{PhysRevB.96.075422}. Negative compressibility has been recently observed in atomically thin BP wherein strong correlations results in an enhanced gate capacitance\cite{PhysRevB.93.035455}. Importantly, negative compressibility occurs at densities as high as $ n \approx 10 ^{12} \text{cm}^{-2}$ which is achievable in experiment. It was shown that an increase in the gate capacitance of a phosphorene field-effect transistor (FET) originates from such negative compressibility at low electron densities\cite{PhysRevB.93.035455}. 

The CDW phase could be identified using scanning tunneling microscopy (STM) via the opening of a gap in the Fermi surface which modifies the local density of states\cite{RevModPhys.60.1129, PhysRevB.37.6571, Coleman:CDW1988, PhysRevLett.104.256403}.  Such a CDW has been also predicted in graphene/\BN/graphene heterostructures in a high magnetic field \cite{PhysRevB.90.035404}. The experimental consequence of a CDW state in such systems would be the reentrant integer quantum Hall effect behavior of electrons in a perpendicular magnetic field. Recently, a system of two strongly coupled electron-hole bilayer graphene sheets was investigated in Ref. [\onlinecite{zarenia2017cdw}], and a new inhomogeneous coupled Wigner crystal phase (c-WC) and  one dimensional (1D) CDW phases were predicted which interplay with the predicted electron-hole superfluid\cite{zarenia2014enhancement,PhysRevB.97.174503}. In such systems, the holes from one layer could behave as a perfectly symmetric polarizable background for the electrons in the other layer, and vice versa. Moreover, the amplitude of the CDW can be tuned by the applied drag force, which depends on the separation between the layers and on temperature\cite{PhysRevB.96.075422}. In graphene, the compressibility does not become negative in the absence of magnetic field because of the positive contribution of the completely filled valence band\cite{RevModPhys.84.1067}. In contrast to graphene, the large gap in phosphorene allows the contribution of the conduction band to dominate and to change the sign of the compressibility of the electron/hole gases at experimentally achievable carrier concentrations.  

\begin{figure}
		\includegraphics[width=9.cm]{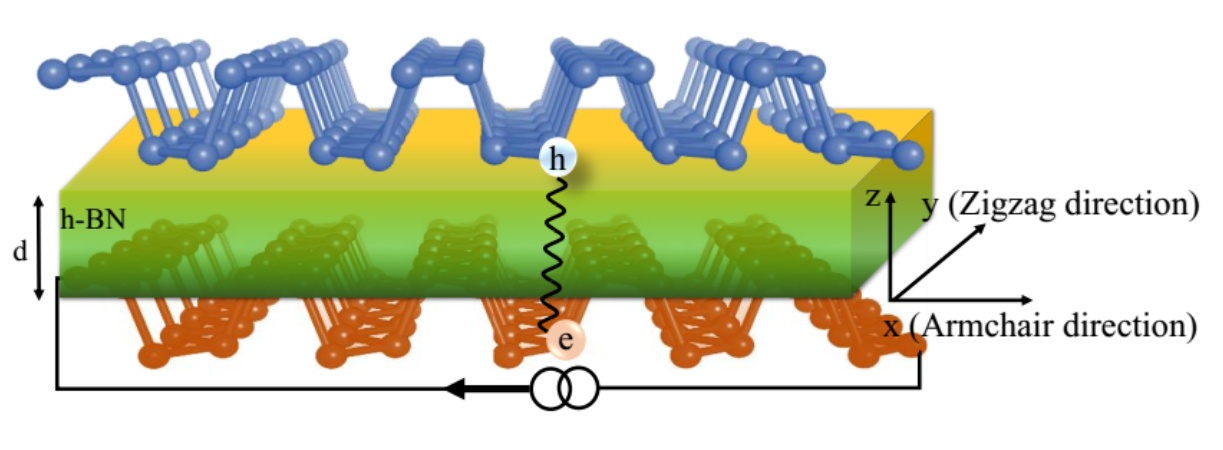}
		\caption {Schematic setup of the electron-hole phosphorene monolayers.}
		\label{Fig1}
\end{figure}

Motivated by the emergence of this experimentally accessible system, we investigate the CDW phase of carriers in coupled anisotropic electron-hole sheets where the electrons and holes interact via the Coulomb interaction. The anisotropy of the charge carriers is modeled by using anisotropic electron and hole masses. In this system, a negative electronic compressibility at sufficiently low density enables the formation of a CDW phase through the application of a uniform force field via the Coulomb drag. In the Coulomb drag setup shown in Fig. \ref{Fig1}, applying the current in one of the layers, i.e. the “active” drive layer, induces an electric field in the other layer, i.e. the “passive” drag layer, in which no current flows. When the compressibility of the drag layer is negative, the force exerted by this electric field results  in CDW phases with a wavelength determined by the absolute value of the compressibility\cite{PhysRevB.96.075422}. We generalize the method of Ref. [\onlinecite{PhysRevB.96.075422}] to investigate the controlled formation of a CDW phase caused by a negative electronic compressibility. The CDW is normally understood to be an intrinsic instability of the electronic structure of the system. The electron liquid is generally protected against such instabilities by the same charge background but at low densities or at high magnetic fields a negative compressibility would be a sign of instability, leading to collapse or a phase transition such as the Wigner crystal \cite{PhysRev.46.1002} or stripe phases \cite{stribephases}. In our proposed electron-hole (e-h) system, the negative compressibility of the layers produce CDW instability when an uniform drag force is applied to the layers. The approach
outlined in this paper is general and will be easily adaptable to other emerging anisotropic 2D materials. Our numerical analytics will be applied to phosphorene.
\\
 The paper is organized as follows. In Sec. \ref{theory}, we obtain the equilibrium solutions for the drag layer modeled as a quasi-1D system of a finite strip of length $L$ with density $n$. We also calculate the anisotropic drag force applied to the drag layer that drives a CDW. We discuss the results of our work in Sec. \ref{results}. Finally, we summarize our results in Sec. \ref{conclusion} and discuss possible experimental observations and consequences.

\maketitle
\section{Theory} \label{theory}
\subsection{Charge density wave in double-layer phosphorene}

We consider two strongly coupled electron and hole layers that are separated by an insulator. We assume that a uniform and steady force is applied to the carriers in each layer due to the momentum transfer processes in a drag set up configuration. We can characterize this force using an interlayer scattering time tensor $\hat{\tau}$, i.e. the rate at which momentum is transfered from the drive layer to  the drag layer. Therefore, the force $\boldsymbol{F}_{d}$  applied from the drive to the drag layer can be written as $\boldsymbol{F}_{d}= \hat{M}\hat{\tau}^{-1} \boldsymbol{v}$ \cite{PhysRevLett.66.1216}, where $\boldsymbol{v}$ is the group velocity of carriers in the drive layer and $\hat{M}$ is the mass tensor with diagonal elements $m^{e/h}_x$ and  $m^{e/h}_y$ along $x$ (armchair) and $y$ (zigzag) directions, respectively. We solve the equilibrium solution for a single layer to which the drag force is applied. The drag layer is modeled as a quasi-1D electron(hole) gas on a finite strip of length $L$ with density $n_{e/h}$. As no current can flow in the direction of the electric field, the drag force must be exactly balanced by the quantum mechanical force arising from the gradient of the chemical potential, $\boldsymbol{F_q=- \nabla} \mu$. We note that $\mu$ is the chemical potential of the interacting system and is related to the ground state energy by $\mu=\partial E/ \partial N$, i.e. $\text{N}$ is the number of particles\cite{Giulianibook,capacitance}. Assuming that the density remains uniform in the zigzag (armchair) direction, the equilibrium condition for the armchair ($x$) direction can be written as 

\begin{equation}
\boldsymbol{F}^{x}_{d}- \frac{\partial \mu}{\partial  x}+\int_{-L_{x}/2}^{L_{x}/2} dx^{\prime} \frac{e^2 \delta n(x^{\prime})}{2 \pi \kappa (x-x^{\prime})}=0,
\label{eq1}
\end{equation}

\noindent where $\delta n$ is the deviation of the 2D electron density from equilibrium with $\kappa$ being the dielectric constant. In the linear regime, the gradient of the chemical potential is

\begin{equation}
\boldsymbol{F}_{q}=- \frac{\partial \mu}{\partial n} \bigg\vert_n  \nabla \delta n.
\end{equation}

Inserting both the drag and quantum mechanical forces in Eq. (\ref{eq1}), we find the equilibrium condition for the $x$ direction 

\begin{equation}
\frac{m^e_{x} v^e_{x}}{\tau_{d}^{x}}-\frac{\partial \mu}{\partial n} \frac{d \delta n (x)}{dx} +\int_{-L/2}^{L/2} dx^{\prime} \frac{e^2 \delta n(x^{\prime})}{2 \pi \kappa (x-x^{\prime})}=0,
\end{equation}

\noindent where $\tau_{d}^{x}$ is the scattering lifetime, the electric field is applied along the $x$ direction in the drive layer and the current is measured along the $x$ direction in the drag layer. We assume $\text{L}_x=\text{L}_y=\text{L}$ and express $x$ and $y$ in units of $\text{L}/2$ and divide the equilibrium solutions by the drag force. Therefore, the equilibrium condition for the $x$ direction of the layer becomes

\begin{equation}
1- \bar{\Gamma}_{x} \frac{d \delta n(x)}{dx}+ (\frac{e^2 \tau^{x}_{d}}{2 \pi \kappa m_{x}^e v_{x}^e})\int_{-1}^{1} dx^{\prime} \frac{\delta n(x^{\prime})}{(x-x^{\prime})}=0,
\label{eq:x}
\end{equation}

\noindent where $\bar{\Gamma}_{x}=\Gamma_{x}/\text{L}$, and $\Gamma_{x}$ is a direction-dependent compressibility-related length given by

\begin{equation}
\Gamma_{x}=\frac{2 \ \tau^{x}_{d}}{v_{x}^e m_{x}^e} \frac{\partial \mu}{\partial n}.
\label{lambdaxy}
\end{equation}

\noindent The density can be expanded in a series of Chebyshev polynomials 

\begin{equation} 
\begin{aligned}
\delta n(x) = \sum_{j=1}^{\infty} c^{x}_j T_{2j-1} (x), \\
\end{aligned}
\end{equation}
\noindent with $c^{x}_j$ the expansion coefficients. Using $ T^{\prime}_n(x)=n U_{n-1}(x)$,  where $U_n(x)$ is associated with the second kind of Chebyshev polynomial, the derivative of the density is given by
\begin{equation}
\begin{aligned}
\frac{d \delta n(x)}{dx}= \sum_{j=1}^{\infty} c^{x}_j (2j-1) U_{2j-2}(x). \\
\label{eq:10}
\end{aligned}
\end{equation}
\\

Equations. (\ref{eq:x})-(\ref{eq:10}) can be solved numerically for $\delta n(x)$. By substituting these expressions into Eq. (\ref{eq:x}), multiplying both sides by $(1-x^2)^{1/2} U_{2k-2}(x)$, and integrating over $x$ with the help of standard integrals for the Chebyshev polynomials one can arrive at a set of linear algebraic equations for the coefficients $c^{x}_j$,

\begin{equation}
\sum_{j=1}^{\infty} W^{x}_{kj}c^{x}_j=\delta_{k1},   \ \ \ \ \  k=1,2...
\label{eq:11}
\end{equation}

\noindent where $W$ is a matrix whose elements are given by

\begin{widetext}
\begin{equation}
W^{x}_{kj}=(\frac{2 \ e^2 \tau_{x}}{ \pi \kappa  m_{x} v_{x}})\big[ \frac{1}{1-4(k+j-1)^2} +   \frac{1}{1-4(k-j)^2}\big] + \bar{\Gamma}_{x}(2k-1) \delta_{kj}.
\label{eq:12}
\end{equation}
\end{widetext}

Using Eqs. (\ref{eq:11}) and (\ref{eq:12}), we find from  Eq. (\ref{eq:x}) the solution

\begin{equation}
\begin{aligned}
\delta n(x)=\sum_{j=1}^{\infty}\ [W^{-1}]^{x}_{j1} \ T_{2j-1}(x),\\
\end{aligned}
\label{eq:13}
\end{equation}

\noindent where $W^{-1}$ is the inverse of the matrix $W$. All equations  apply to the zigzag ($y$) direction. In the next section, we obtain the interlayer scattering time.

 \subsection{Interlayer Scattering Rate }
 
 The Coulomb drag technique allows a unique access to the interlayer scattering rate via a resistance measurement. This can be simply shown within the Drude model \cite{RevModPhys.88.025003}. Considering an electron-hole double-layer system, the anisotropic drag resistivity is given by \cite{Samira:drag2016}
 
 \begin{equation}
 \begin{aligned}
 \rho_{d}^{\alpha \beta}=&\frac{\mathrm{\ensuremath{\hbar^2}}}{2\pi e^{2}n_{1}n_{2}k_{B}T}\int\frac{d^{2}q}{(2\pi)^{2}}q_{\alpha}q_{\beta}\\   
 &\times\int_{0}^{\infty}d\omega\frac{|U_{eh}(\boldsymbol{q},\omega)|^{2} \text{Im}\Pi_{h}(\boldsymbol{q},\omega) \text{Im}{\Pi_{e}(\boldsymbol{q},\omega)}}{\sinh^{2}(\hbar \omega/2k_{B}T)},
 \label{eq:drag}
 \end{aligned}
 \end{equation}
 
 \noindent where the $\alpha$ and $\beta$ indices refer to the $x$ and $y$ components. We assume that the drive layer contains electrons and the drag layer contains holes. Therefore, the drag resistivity depends on the interlayer momentum relaxation rate as 
 
 \begin{equation}
\rho_{d}^{\alpha \beta}= \frac{\hat{M}^e_{\alpha}}{n_e e^2 \tau^{\alpha \beta}_{d}}.
\label{eq:tau}
 \end{equation}
 
  The dynamically screened interlayer potential $U_{eh}(\boldsymbol{q},\omega)$ can be obtained by solving the corresponding Dyson equation \cite{Badalyan:prb12},
 \begin{equation}
 U_{eh}(\boldsymbol{q},\omega)=\frac{V_{eh}(q)}{\det| \varepsilon_{eh}(\boldsymbol{q},\omega)|},
 \label{eq:Ueh}
 \end{equation}
 where $V_{eh}(q)=\nu(q)\exp(-qd)$ is the unscreened 2D Coulomb interaction, with $d$ being the interlayer separation. $\nu(q)=-2\pi e^{2}/q \kappa$ is the Coulomb potential and $\varepsilon_{eh}(\boldsymbol{q},\omega)$ is the dynamic dielectric matrix of the system. Using the random phase approximation (RPA) formalism, we have\cite{Hwang:prb09,Wen:nanotech12}
 \begin{equation}
 \varepsilon_{eh}(\boldsymbol{q},\omega)=\delta_{eh}+V_{eh}(q)\Pi_{e}(\boldsymbol{q},\omega).
 \label{eq:epsil}
 \end{equation}
 \noindent For an electron gas system, the noninteracting density–density response function can be obtained from the following equation\cite{Bohm:pr53}:
 \begin{equation}
 \Pi_{e/h}(\boldsymbol{q},\omega)=-\frac{g_{s}}{\nu} \sum \limits_{\boldsymbol{k}} \frac{f^{0}(E_{\boldsymbol{k}}^{e/h})-f^{0}(E_{\boldsymbol{k+q}}^{e/h})}{E_{\boldsymbol{k}}^{e/h}-E_{\boldsymbol{k+q}}^{e/h}+\hbar\omega+i\eta}.
 \label{eq6}
 \end{equation}
 Here $f^{0}(E_{\boldsymbol{k}}^{e/h})$ is the Fermi-Dirac distribution, $g_{s}=2$ is the spin degeneracy, $\nu$ is the unit cell surface, and $\eta$ is the broadening parameter that may account for disorder in the system. Here, we extend our model to the case of e-h sheets where the energy bands are anisotropic. The corresponding Hamiltonian point can be expressed as \cite{Rodin:prl2014}
  
  \begin{equation}
  \hat{H}_{\textbf{0}}=\left(\begin{array}{cc}
  E_c + \eta_c k^2_x + \alpha_c k^2_y 
  & \gamma k_x+\beta k_y^2\\
  \gamma k_x+\beta k_y^2 &E_v-\eta_v k^2_x - \alpha_v k^2_y 
  \end{array}\right),
  \label{eqH}
  \end{equation}
  
  \noindent where  $E_{c}$ ($E_{v}$) is the energy of conduction (valence) band edge, and $\gamma$ and $\beta$ describes the effective coupling between the conduction and valence bands. $\eta_{c,v}$ and $\alpha_{c,v}$ are related to the anisotropic effective masses, which is our numerical analysis and will be taken like that of phosphorene\cite{Rodin:prl2014},
  
  \begin{equation}
  \begin{aligned}
  &m^x_{e}=\dfrac{{\hbar^2}}{2(\eta_c+\gamma^2/2E_g)},\\
  &m^x_{h}=\dfrac{{\hbar^2}}{2(\eta_v-\gamma^2/2E_g)},\\
  &m^y_{e(h)}=\dfrac{{\hbar^2}}{2(\alpha_{c(v)})},\\
  \end{aligned}\label{16}
  \end{equation} 
  
   \noindent where $E_g$ is the energy band gap. Using Eq. ({\ref{16}}), one can then use these masses to obtain an approximation for the spectrum \cite{Low:prl14,PhysRevB.93.165402},

 \begin{equation}
 E_{\boldsymbol{k}}^{e/h}=\frac{\hbar^2}{2}\big(\frac{k_{x}^{2}}{m^{x}_{e/h}}+\frac{k_{y}^{2}}{m^{y}_{e/h}}\big)-\mu_{e/h},
 \label{eq:energy}
 \end{equation}
 
\noindent where $\mu_{e/h}$ is the
chemical potential of the electron(hole) layers. The temperature-dependent dynamical density-density response function for intra-band transitions can be calculated

 \begin{equation}
 \begin{aligned}
& \frac{\Pi_{e/h}(\boldsymbol{q},\omega)}{g^{e/h}_{2D}}=-\int dK \frac{\Phi_{e/h}(K,T)}{Q_{e/h}}  \\
 &\times\Bigg[\text{sgn}(\text{Re}(Z_{-})) 
 \frac{1}{\sqrt{Z_{-}^{2}-K^{2}}}
 -sgn(\text{Re}(Z_{+}))\frac{1}{\sqrt{Z_{+}^{2}-K^{2}}}\Bigg].
 \end{aligned}
 \end{equation}
 In the above symmetric form of temperature-dependent anisotropic density-density response function we have defined $\boldsymbol{Q_{e/h}}=(\boldsymbol{q}/k^{e/h}_{F})\sqrt{m^{e/h}_D/ \hat{M}^{e/h}} $, $\boldsymbol{K_{e/h}}=(\boldsymbol{k}/k^{e/h}_{F})\sqrt{m^{e/h}_D/\hat{M}^{e/h}}$, $g^{2D}_{e/h}=m^{e/h}_D / \pi \hbar^{2}$, $Z_{e/h}^{\pm}=((\hbar\omega+i\eta)/\hbar k^{e/h}_F Q_{e/h} \nu^{e/h}_{F})\pm(Q_{e/h}/2)$ with $\nu^{e/h}_{F}=\hbar k^{e/h}_{F}/m^{e/h}_D$, $m^{e/h}_D=\sqrt{m_x^{e/h}m_y^{e/h}}$ is the 2D density of state mass, and 
 
 \begin{equation}
 \Phi_{e/h}(K_{e/h},T)=\frac{K_{e/h}}{1+\exp[(K_{e/h}^{2}E_{F}^{e/h}-\mu^{e/h})/k_{B}T]},
 \end{equation}
 
\noindent where $k^{e/h}_{F}=\sqrt{2\pi n^{e/h}}$ and $\mu_{e/h}$ is the chemical potential of the electron and hole layers which is determined by the particle number conservation condition \cite{Ashcroft}. 
We work in polar coordinates $\boldsymbol{q}=q(\cos\theta,\sin\theta)$,  $Q_{e/h}=(q/k^{e/h}_{F})\sqrt{m^{e/h}_D R^{e/h}(\theta)}$ in which the orientation factor $R_{e/h}$ is expressed as $
 R^{e/h}(\theta)=\big(\cos^{2}(\theta)/m_x^{e/h}+\sin^{2}(\theta)/m_y^{e/h}\big)$. At sufficiently low temperature and  $q \ll k^{e/h}_{F}$ one can approximate $\text{Im}[\Pi_{e/h}(\boldsymbol{q},\omega)]$ by the low frequency expression

\begin{equation}
\text{Im}[{\Pi_{e/h}(\boldsymbol{q},\omega)}] \approx - \frac{(m^{e/h}_D)^{2} \ \omega}{2 \pi \hbar^3 Q_{e/h} (k_F^{e/h})^2}.
\label{24}
\end{equation}

 In this case, the screened potential of the electron-hole interaction in Eq. (\ref{eq:Ueh}) $ |U_{eh}(\boldsymbol{q},\omega)|$ can be approximated by the static interlayer Coulomb interaction in the long wave-length limit as

\begin{equation}
U_{eh}(q \to 0)= \frac{v_q e^{-qd}}{[1-v_q  \frac{\partial n_{e}}{\partial \mu_{e}}] [1-v_q  \frac{\partial n_{h}}{\partial \mu_{h}}] - \frac{\partial n_{e}}{\partial \mu_{e}} \frac{\partial n_{h}}{\partial \mu_{h}}  v_q^2 e^{-2qd}}.
\label{eq:23}
\end{equation}

	\begin{figure*}[ht]
	\includegraphics[width=18.cm]{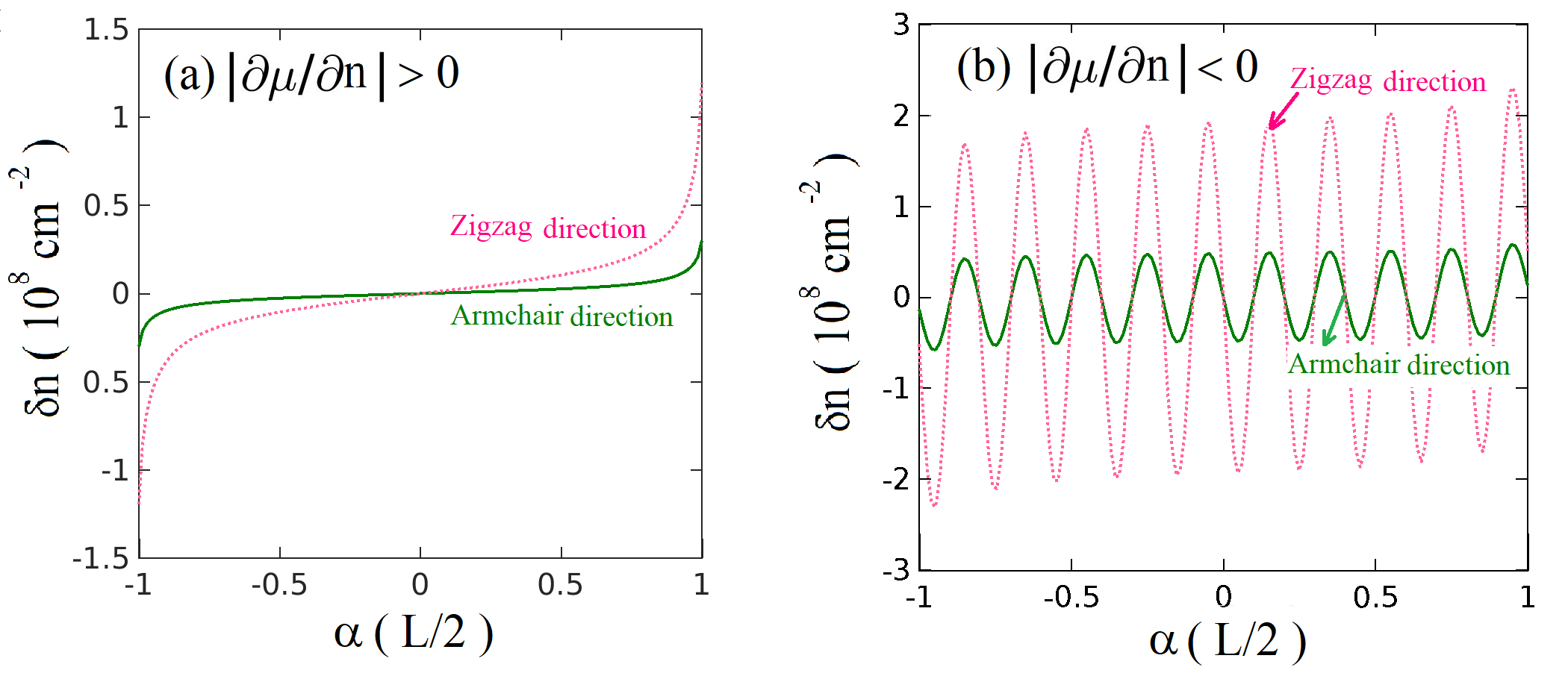}
	\caption {The equilibrium solutions of Eq. (\ref{eq:x}) for (a) a positive compressibility ($\partial \mu / \partial n > 0$) and (b) a negative compressibility ($\partial \mu / \partial n < 0$) with $ |\partial \mu / \partial n|\approx $  5 $ \times 10^{-12}$ \text{meV cm$^{2}$} along the armchair(x) and zigzag(y) directions at the equilibrium density $n= 2 \times 10 ^{12} cm^{-2}$. Here $d=2 nm$ and $T=2K$. The length of the strip is taken to be L = 5$\lambda$. Other parameters were taken from phosphorene.}
	\label{Fig2}
\end{figure*}

\noindent An important property of this equation is the compressibility, related to the small $q$ limit of the proper density-density response function of the compressibility
sum rule \cite{Giulianibook}. There is an exact relation between the compressibility and the long wave-length limit of the static density-density response function as $\Pi_{e/h}(q \to 0)=-\partial n_{e/h}/ \partial \mu_{e/h}$\cite{Giulianibook}. Inserting Eq. (\ref{eq:23}) in Eq. (\ref{eq:drag}) we have

\begin{equation}	
U_{eh}(q \to 0)=\frac{ -2\pi e^2}{ \kappa} \frac{ q}{\epsilon_s(q)},
\label{eq:U} 
\end{equation}

\noindent where $\epsilon_s(q)$ is defined as

\begin{equation}
\epsilon_s(q)=[q^2 e^{qd}+q(q^e_s+q^h_s)e^{qd}]+q^e_s q^h_s (e^{qd}-e^{-qd}),
\label{es}
\end{equation}

 \noindent here $q^{(e/h)}_s= (2 \pi e^2/ \kappa)( \partial n_{e/h} /\partial \mu_{e/h})$ is the screening wave vector related to compressibility. Since the most important contribution from the integration over $q$ comes from the region in which $q \le 1/d$, we may neglect the first term in the denominator of Eq. (\ref{eq:U}) and obtain the simple form 

	\begin{equation}	
	U_{eh}(q \to 0)=\frac{ -2\pi e^2}{\kappa} \frac{ q}{ 2 q^e_{s} q^h_{s} \sinh(qd)}.
	\label{eq:U2} 
	\end{equation}

Using Eq. (\ref{eq:U2}), the drag resistivity in Eq. (\ref{eq:drag}) is given by

\begin{widetext}
	\begin{equation}
	\rho_{eh}^{\alpha \beta} \approx  \frac{\hbar^2}{(2\pi)^3 e^{2}n_{h}n_{e}k_{B}T} \frac{(m^e_D)^2 (m^h_D)^2}{(2 \pi \hbar)^3 k^e_F k^h_F} \int d\theta \int \frac{q\ dq\ q^{\alpha}_e \ q^{\beta}_h }{Q_e Q_h} (\frac{\pi e^2}{\kappa})^2 \frac{q^2}{(q_{de} q_{dh})^2(\sinh^2(qd))}  
	\int_{0}^{\infty}d\omega \frac{ \omega^2}{\sinh^2(\frac{\hbar \omega}{(2k_B T)}}.
		\label{eq:drag2}
		\end{equation}
	\end{widetext}
		
This allows one to analytically obtain the integration over $q$ and $\omega$, by using $\int_{0}^{\infty} dx x^p/4 \sinh^2(x/2) = p! \ \zeta(p)$\cite{tableintegral}. Since $	\rho_{eh}^{xy} \propto \int  d\theta \sin(\theta)  \cos(\theta)=0 $, the off diagonal elements of the drag resistivity vanish. The analytical results for the diagonal elements of the drag resistivity at low temperatures ($T\ll T_F$) becomes 

\begin{equation}
	\rho^{\alpha \alpha}_{eh}= \frac{{e^2} \zeta(3) } {16 \pi^3 \hbar^2} \frac{(k_B T)^2 \sqrt {m^e_{\alpha} m^h_{\alpha} m^e_d m^h_d} }{n_e n_h E^e_F E^h_F} \frac{1}{(q^e_sd)^2 (q^h_sd)^2}.
	\label{eq25}
\end{equation}

One can then use this equation into Eq. (\ref{eq:tau}) to get a diagonal tensor for the anisotropic interlayer scattering rate 

\begin{equation}
\tau^{\alpha \alpha}_{d}=  \frac{16 \pi^3 \hbar^2 (q^e_sd)^2 (q^h_sd)^2 }{{e^4} \zeta(3) }  \frac{n_h E^e_F E^h_F \ \hat{M}^{\alpha}_e }{(k_B T)^2 \sqrt {m^{\alpha}_e m^{\alpha}_h m^e_D m^h_D} }.
\end{equation}

For a system with equal electron and hole densities $n_{e}=n_{h}=n$ and equal chemical potentials, we have $q^e_s=q^h_s = (8 \pi^2 / \lambda)$, where $\lambda=(4 \pi \kappa / e^2) ( \partial \mu / \partial n)$ is the compressibility related length \cite{PhysRevB.96.075422}.

 \section{Results and discussion} \label{results}

\begin{figure}[t]
	\includegraphics[width=9.cm]{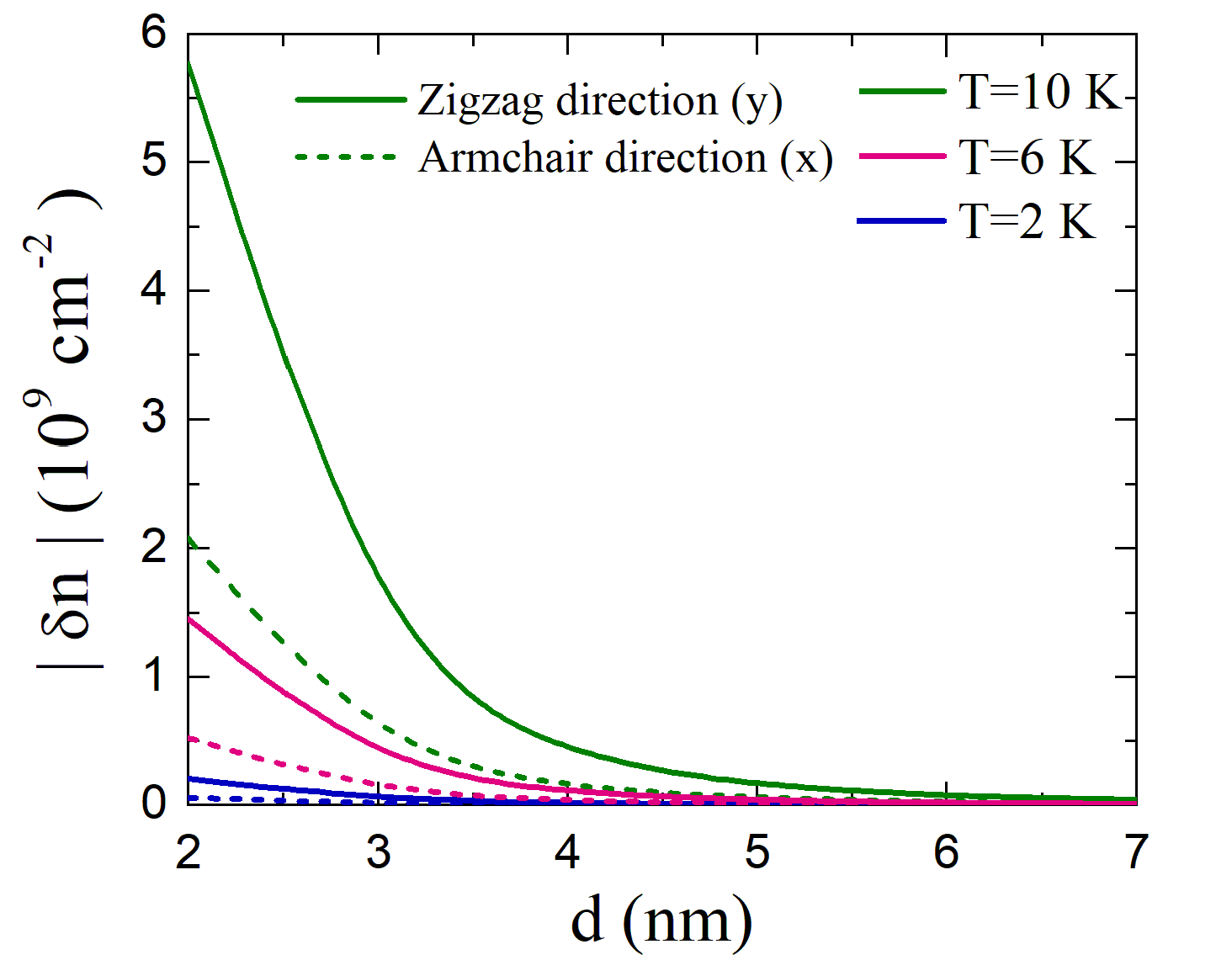}
	\caption {Amplitude of the charge modulation $|\delta n| $ for a negative compressibility with $ \partial \mu / \partial n \approx $ - 5 $ \times 10^{-12}$ \text{meV cm$^{2}$} along the zigzag ($y$) and armchair ($x$) directions as a function of the separation between the BP layers $d$ at different temperatures for equilibrium density $n= 2 \times 10 ^{12} cm^{-2}$ and L = 5$\lambda$.
	}
	\label{Fig3}
\end{figure}

 The density modulation depends directly on the system anisotropy and also on the sign of the compressibility. In order to explore the CDW in the phosphorene layer subjected to the drag force, we study the effects of the sign of the compressibility on the equilibrium solution, i.e. Eq. (\ref{eq:x}). In Fig. \ref{Fig2}(a) we show the numerical results as calculated from the equilibrium solution for a positive compressibility ($\partial \mu / \partial n  > 0$ ) in the armchair ($x$) and zigzag ($y$) directions of phosphorene with effective masses $m^e_x = 0.16 m_0$ and $m^e_y =1.24 m_0$ and $m^h_x = 0.15m_0$ and $m^h_y =4.95m_0$ with the velocity $v^e_{x} = 6 \times 10^6 \ (cm/s)$ and $v^e_{y}= 1.5 \times 10^6 \ (cm/s)$\cite{Zhenghe:apl16,PhysRevB.90.085402} with $m_0$ being the free electron mass. We set the dielectric constant $\kappa \approx 4$ \cite{constantinescu2016multipurpose} for encapsulated phosphorene by \BN \ layers\cite{PhysRevB.96.075422} and L$=5 \lambda$. We can see that similar to the case of an isotropic 2DEG \cite{PhysRevB.96.075422} the CDW is not stable for phosphorene along the directions where the compressibility is positive.
 Note that, the charge does not change at the center of phosphorene but it accumulates along the edges due to the strong electric field induced at the sharp edges. As a direct consequence of the band anisotropy the charge accumulation is different at the two directions of phosphorene. 
 We show the charge modulation for phosphorene layers in Fig. \ref{Fig2}(b) when the compressibility is negative. One can see that the density modulations along the zigzag ($y$) and armchair ($x$) directions have different amplitudes. The amplitude and frequency of these oscillations depend on the applied electric field and the value of $\Gamma_{\alpha}$ in our Coulomb drag setup providing opportunity for electronic tuning. Moreover, phosphorene shows CDWs at higher densities as compared to typical 2D electron gas systems \cite{PhysRevB.96.075422}. It is interesting to note that the periodicity of the CDW can be electrically tuned, via a gate, by the density of the drag layer, and hence the value of $|\lambda|$.  As one can also see from Fig. \ref{Fig2}(b), the wavelength of the CDW which is $|\lambda|/2 $ does not depend on the direction. The Coulomb interaction between the holes and electrons can also significantly enhance the occurrence of the CDW phase, especially when $r_s \gg d$, where $r_s=1/(a^{*(e/h)}_B\sqrt{n \pi})$ is the average distance between two neighboring in-plane carriers and $a^{* (e/h)}_B= (\hbar \kappa / m^{e/h}_D e^2)$, the effective Bohr radius. Figure. \ref{Fig3} shows the amplitude of the CDW along both zigzag and armchair directions as a function of the separation between the two layers at different temperatures. We find that the CDW state does become stable as the spacing between the layers is decreased. When we increase the separation $d$, the density modulation decreases because at sufficiently large $d$, the strength of the interlayer attraction is decreased. Moreover, as expected for Coulomb scattering in Fermi systems, the drag resistivity is approximately proportional to $T^2$ (see Eq. (\ref{eq25}) in the manuscript and also Ref. [\onlinecite{PhysRevLett.102.026804}]). Therefore, higher temperatures result in larger drag and thus larger density modulation amplitudes. Also, it is predicted that the drag resistance will increase dramatically at the critical temperature below $1$ K where the paired electrons and holes want to move together to form an exciton condensate which is beyond our approach \cite{PhysRevLett.102.026804,Malliakas}. At very high temperatures, $T\gg T_F$, the thermal fluctuations destroy any formation of a CDW phase. However, the results in Fig. \ref{Fig3} are at temperatures with $T\ll T_F$ (the equilibrium density $n= 3 \times 10 ^ {12} cm^{-2}$ corresponds to $T_F \approx 170$ K) where carrier backscattering is neglected which will destroy the coherence \cite{PhysRevLett.102.026804,PhysRevB.78.075430}.

\section{conclusion}  \label{conclusion}
 In this paper, we considered a system of two strongly coupled electron-hole layers interacting by the Coulomb interaction. The system we proposed for phosphorene is expected to support large transresistivity \cite{Samira:drag2016} due to the large effective mass of the carriers and consequently a large CDW modulation. Electrons and holes in the device are induced with top and bottom gates leading to controllable density in each layer.  Separate electrical contacts to the two electron and hole layers allow the injection of current into one and the detection of a small drag voltage across the other. The drag voltage is a direct measure of the inter layer momentum relaxation rate. Measurements of this rate are in qualitative agreement with calculations of an interlayer Coulomb scattering model.
 	
  As we have mentioned, for an ordinary system the very same charge background that allows the compressibility to go negative is a major obstacle to the formation of charge-density waves due to the electrostatic energy cost that such waves would incur. However, in our manuscript and also in Ref. [\onlinecite{PhysRevB.96.075422}], it was shown that a uniform and steady force applied to the carriers when the compressibility is negative produces a charge-density wave. The force arising from the Coulomb drag effect is one way to apply such a steady force. Moreover, recent experiments on graphene-terminated BP heterostructures show an enhanced capacitance $C_t=C_g/(1+1/e^2 d\mu/dn$),i.e., much larger than the expected geometric capacitance $C_g=\kappa/d$ \cite{PhysRevB.93.035455}. This enhanced capacitance results from a negative quantum capacitance $C_q=e^2  dn/d\mu$ when the compressibility ($d\mu/dn<0$ ) becomes negative. We have shown that a negative compressibility at low carrier concentrations by using the external force that arises from the Coulomb drag effect is one way to produce the density modulation in the drag layer. We obtained the equilibrium solutions for the drag layer modeled as a quasi-1D system. By using the compressibility sum rule in which the compressibility is related to the small momentum limit of the proper density-density response function, an analytical expression was obtained for the anisotropic drag force. We found that the density modulation not only depends on the sign of the compressibility but also depends on the system anisotropy. The amplitude of
the CDW phase depends on the separation between the layers and temperature. The wave-length of the CDW can be also electrically tuned by changing the value of $\Gamma_{x}$. This might be changed in experiment , via a gate, changing the density of the passive layer. 

There are a number of ways that a CDW phase could be experimentally identified \cite{PhysRevB.52.14516}. This phase can be detected by optical methods, such as differential absorption or diffraction and also by using STM, which can detect a number of features related to the onset of CDW, including charge modulation, periodic distortion of atomic position, and the opening of a gap in the density of states. Angle resolved photoemission spectroscopy (ARPES) can also show which portions of the Fermi surface are gapped out by the CDW \cite{PhysRevLett.99.046401}. This can give guidance as to whether the CDW is driven by Fermi surface nesting or by something more exotic \cite{1367-2630-10-5-053019}. By perturbing
a material with an ultrashort laser pulse and following the resulting transient ARPES spectra, one can obtain insight into the dynamics both of quasiparticle occupations and of the electronic structure itself \cite{0295-5075-115-2-27001}. To evaluate the feasibility of optical detection of such an anisotropic CDW in BP, we note that the angular dependence of the differential reflection signal can be induced by both the pump absorption and the probe detection efficiency \cite{PhotocarriersBP:nanol15}. It was shown that by rotating the sample with respect to the pump polarization, the pump injects different carrier densities and the anisotropic band structure causes the probe pulses with different polarizations to sense the carriers with different efficiencies \cite{PhotocarriersBP:nanol15}.  The CDW phase could also be identified using STM. Due to the formation of the CDW along with the associated periodic lattice distortion, a gap in the Fermi surface opens. This modifies the local density of states which could be detected by tunneling \cite{RevModPhys.60.1129, PhysRevB.37.6571, Coleman:CDW1988, PhysRevLett.104.256403}. A CDW phase can be also detected by transport measurements. 1D stripes may be pinned by disorder, in which case the CDW could be identified either with standard threshold voltage conductivity measurements or with frequency threshold measurements of ac conductivity \cite{PhysRevB.38.13019}.  Our approach is general and applicable to other 2D materials with anisotropic parabolic energy bands. 
\\
\section*{Acknowledgements}
This work was partially supported by the Flemish Science Foundation (FWO-Vl) and the Methusalem program of the Flemish government and Iran Science Elites Federation.
                   
%

\end{document}